\documentclass[useAMS,usenatbib]{mn2e}
\usepackage{graphicx}
\usepackage{times}
\usepackage{mn2e-breakabs}
\usepackage{amssymb}

\begin{document}

\title[Non-Gaussianity in BOSS 3D Clustering] 
{The Clustering of Galaxies in SDSS-III DR9 Baryon Oscillation Spectroscopic Survey: Constraints on Primordial Non-Gaussianity}

\author[A. J. Ross et al.]{\parbox{\textwidth}{
Ashley J. Ross\thanks{Email: Ashley.Ross@port.ac.uk}$^{1}$, 
Will J. Percival$^{1}$,
Aurelio Carnero$^{2,3}$,
Gong-bo Zhao$^1$,
Marc Manera$^1$,
Alvise Raccanelli$^{1,4,5}$,
Eric Aubourg$^6$,
Dmitry Bizyaev$^7$,
Howard Brewington$^7$,
J. Brinkmann$^7$,
Joel R. Brownstein$^8$,
Antonio J. Cuesta$^9$,
Luiz A. N. da Costa$^{2,3}$,
Daniel J. Eisenstein$^{10}$,
Garrett Ebelke$^7$,
Hong Guo$^{11}$,
Jean-Christophe Hamilton$^{12}$,
Mariana Vargas Maga\~na$^{12}$,
Elena Malanushenko$^7$,
Viktor Malanushenko$^7$,
Claudia Maraston$^1$,
Francesco Montesano$^{13}$,
Robert C. Nichol$^1$,
Daniel Oravetz$^7$,
Kaike Pan$^7$,
Francisco Prada$^{14,15}$,
Ariel G. S\'anchez$^{13}$,
Lado Samushia$^{1,16}$,
David J. Schlegel$^{17}$,
Donald P. Schneider$^{18,19}$,
Hee-Jong Seo$^{20}$,
Alaina Sheldon$^7$,
Audrey Simmons$^7$,
Stephanie Snedden$^7$,
Molly E. C. Swanson$^{10}$,
Daniel Thomas$^1$,
Jeremy L. Tinker$^{21}$,
Rita Tojeiro$^1$,
Idit Zehavi$^{11}$}
  \vspace*{4pt} \\ 
$^{1}$Institute of Cosmology \& Gravitation, Dennis Sciama Building, University of Portsmouth, Portsmouth, PO1 3FX, UK\\ 
$^{2}$Observat\'orio Nacional, Rua Gal. Jos\'e Cristino 77, Rio de Janeiro, RJ - 20921-400, Brazil\\
$^3$Laborat\'orio Interinstitucional de e-Astronomia - LineA, Rua Gal. Jos\'e Cristino 77, Rio de Janeiro, RJ - 20921-400, Brazil\\
$^4$Jet Propulsion Laboratory, California Institute of Technology, Pasadena, CA 91109, USA\\
$^5$California Institute of Technology, Pasadena, CA 91125, USA\\
$^6$APC, Univ Paris Diderot, CNRS/IN2P3, CEA/Irfu, Obs de Paris, Sorbonne Paris Cit\'e, France\\
$^7$Apache Point Observatory, P.O. Box 59, Sunspot, NM 88349-0059, USA\\
$^8$Department of Physics and Astronomy, University of Utah, Salt Lake City, UT 84112, USA\\
$^{9}$Yale Center for Astronomy and Astrophysics, Yale University, New Haven, CT 06511, USA\\
$^{10}$Harvard-Smithsonian Center for Astrophysics, 60 Garden St., MS \#20, Cambridge, MA 02138, USA\\
$^{11}$Department of Astronomy, Case Western Reserve University, Cleveland, Ohio 44106, USA\\
$^{12}$APC, Universit\'e Paris-Diderot-Paris 7, CNRS/IN2P3, CEA, Observatoire de Paris, 10, rue A. Domon \& L. Duquet, Paris, France\\
 $^{13}$Max-Planck-Institut f\"ur extraterrestrische Physik, Postfach 1312, Giessenbachstr., 85748 Garching, Germany\\
$^{14}$Campus of International Excellence UAM+CSIC, Cantoblanco, E-28049 Madrid, Spain\\
$^{15}$Instituto de Fisica Teorica (UAM/CSIC), Universidad Autonoma de Madrid, Cantoblanco, E-28049 Madrid, Spain\\
$^{16}$National Abastumani Astrophysical Observatory, Ilia State University, 2A Kazbegi Ave., GE-1060 Tbilisi, Georgia\\
$^{17}$Lawrence Berkeley National Laboratory, 1 Cyclotron Road, Berkeley, CA 94720, USA\\
$^{18}$Department of Astronomy and Astrophysics, The Pennsylvania State University, University Park, PA 16802, USA\\
$^{19}$Institute for Gravitation and the Cosmos, The Pennsylvania State University, University Park, PA 16802, USA\\
$^{20}$Berkeley Center for Cosmological Physics, LBL and Department of Physics, University of California, Berkeley, CA 94720, USA\\
$^{21}$Center for Cosmology and Particle Physics, New York University, New York, NY 10003, USA\\
}

\date{accepted by MNRAS} 
\pagerange{\pageref{firstpage}--\pageref{lastpage}} \pubyear{2012}
\maketitle
\label{firstpage}

\begin{abstract}
We analyze the density field of 264,283 galaxies observed by the Sloan Digital Sky Survey (SDSS)-III Baryon Oscillation Spectroscopic Survey (BOSS) and included in the SDSS data release nine (DR9). In total, the SDSS DR9 BOSS data includes spectroscopic redshifts for over 400,000 galaxies spread over a footprint of more than 3,000 deg$^2$. We measure the power spectrum of these galaxies with redshifts $0.43 < z < 0.7$ in order to constrain the amount of local non-Gaussianity, $f_{\mathrm{NL}}^{\mathrm{local}}$, in the primordial density field, paying particular attention to the impact of systematic uncertainties. The BOSS galaxy density field is systematically affected by the local stellar density and this influences the ability to accurately measure $f_{\mathrm{NL}}^{\mathrm{local}}$. In the absence of any correction, we find (erroneously) that the probability that $f_{\mathrm{NL}}^{\mathrm{local}}$ is greater than zero, $P(f_{\mathrm{NL}}^{\mathrm{local}}>0)$, is 99.5\%. After quantifying and correcting for the systematic bias and including the added uncertainty, we find -45 $< f_{\mathrm{NL}}^{\mathrm{local}} < $ 195 at 95\% confidence, and $P(f_{\mathrm{NL}}^{\mathrm{local}}>0) = 91.0\%$. A more conservative approach assumes that we have only learned the $k$-dependence of the systematic bias and allows any amplitude for the systematic correction; we find that the systematic effect is not fully degenerate with that of $f_{\mathrm{NL}}^{\mathrm{local}}$, and we determine that -82 $< f_{\mathrm{NL}}^{\mathrm{local}} < $178 (at 95\% confidence) and $P(f_{\mathrm{NL}}^{\mathrm{local}}>0) = 68\%$. This analysis demonstrates the importance of accounting for the impact of Galactic foregrounds on $f_{\mathrm{NL}}^{\mathrm{local}}$ measurements. We outline the methods that account for these systematic biases and uncertainties. We expect our methods to yield robust constraints on $f_{\mathrm{NL}}^{\mathrm{local}}$ for both our own and future large-scale-structure investigations.
\end{abstract}

\begin{keywords}
  cosmology: observations - (cosmology:) inflation - (cosmology:) large-scale structure of Universe
\end{keywords}

\section{Introduction}

Measuring the clustering of galaxies on large scales provides an exciting opportunity to test inflationary models. Slow roll, single field inflation is believed to generate a primordial gravitational potential (hereafter potential) that can be described statistically by a (nearly) Gaussian random field \citep{Bardeen86}. Alternative inflationary models (e.g., multiple field) predict there to be significant non-Gaussian components to the potential (see, e.g., \citealt{Wands10} for a review). For a Gaussian random field, the amplitude of fluctuations in 3-point configurations is zero. It is therefore convenient to express the degree of non-Gaussianity, $f_{\mathrm{NL}}$, as a dimensionless ratio between the amplitude of the bispectrum, $B_{\Phi}(k_1,k_2,k_3)$, and power spectrum, $P_{\Phi}(k)$, of the fluctuations of the total potential\footnote{Specifically, Bardeen's gauge-invariant potential, which is equivalent to the Newtonian potential at sub-horizon scales} $\Phi$: 
\begin{equation}
f_{\mathrm{NL}} \equiv \frac{B_{\Phi}(k_1,k_2,k_3)}{2\left[P_{\Phi}(k_1)P_{\Phi}(k_2)+P_{\Phi}(k_2)P_{\Phi}(k_3)+P_{\Phi}(k_1)P_{\Phi}(k_3)\right]}.
\label{eq:fnldef}
\end{equation}

One can denote the portion of the potential that can be described as a Gaussian random field as $\phi$ and assume that $f_{\mathrm{NL}} $ is a function of the potential (i.e., that it is `local'). To 2nd order, this approach yields
\begin{equation}
\Phi = \phi + f_{\mathrm{NL}}^{\mathrm{local}}(\phi^2-\langle\phi^2\rangle).
\end{equation}
For this simplest treatment and evaluating Eq. \ref{eq:fnldef} in the limit where $k$-space triangle configurations satisfy $|\vec{k}_{12}|\ll |\vec{k}_{13}|,|\vec{k}_{23}|$ (known as the ``squeezed'' limit), it has been shown that a bias in the dark matter halo power spectrum proportional to $f_{\mathrm{NL}}^{\mathrm{local}}k^{-2}$ would result (\citealt{Dalal08,Matt08}). \cite{Komatsu10} and \cite{Crem11} have shown that, for standard, single-field inflation, the amplitude of the bispectrum in the ``squeezed'' limit becomes vanishingly small, and thus any detected scale-dependent bias at large-scales would disprove all single-field models, subject to the condition that the field starts in the vacuum. Therefore, measurements of the large-scale clustering of galaxies have the potential to yield significant insight into the physical mechanism that drove inflation.

Primordial non-Gaussianity can also be measured from the bispectrum of cosmic microwave background (CMB) anisotropies (see, e.g., \citealt{Bartolo04,Komatsu10} and references therein) and, in principle, galaxies (see e.g. \citealt{Sef07,Scocc12}). WMAP7 found $-10 < f_{\mathrm{NL}}^{\mathrm{local}} < 74$ to 95\% confidence \citep{KomatsuWMAP7}. To date, the reported constraints on $f_{\mathrm{NL}}^{\mathrm{local}}$ from galaxy power spectrum measurements have been competitive with those obtained from the CMB bispectrum. \cite{Slosar08} used a combination galaxy and quasar clustering measurements to find $-29 < f_{\mathrm{NL}}^{\mathrm{local}} < 70$ at 95\% confidence. \cite{Xia11} analyzed similar measurements based on updated data to decrease the 95\% confidence interval to $5 < f_{\mathrm{NL}}^{\mathrm{local}} <  84$. \cite{KomatsuWMAP7} combined their CMB results with the \cite{Slosar08} result to obtain $-5 < f_{\mathrm{NL}}^{\mathrm{local}} < 59$ at 95\% confidence. Current results thus favour positive $f_{\mathrm{NL}}^{\mathrm{local}}$, but do not rule out $f_{\mathrm{NL}}^{\mathrm{local}} = 0$. \cite{Giann12} predict that combining CMB data from the {\it Planck} mission \citep{Planck} and galaxy data from a Euclid-like \citep{euclid} survey will reduce the 1$\sigma$ uncertainty on $f_{\mathrm{NL}}^{\mathrm{local}}$ to 3.

A large value of $f_{\mathrm{NL}}^{\mathrm{local}}$ implies larger amplitudes in the 2-point clustering of galaxies at large separations than expected in the standard $\Lambda$CDM paradigm. Recent studies (e.g., \citealt{Thomas11,Saw11,Nik12}) have indeed found larger than expected clustering amplitudes at large scales, using photometric Sloan Digital Sky Survey (SDSS; \citealt{SDSS}) data. However, \cite{imsys} showed that the excess found in \cite{Thomas11} was due, at least partially, to stellar contamination, and that systematics correlated with the Galaxy (e.g., stellar density and Galactic extinction) may impart spurious clustering signal at large scales. \cite{Nik12} find their measurements yield $f_{\mathrm{NL}}^{\mathrm{local}} = 90\pm30$ at 68\% confidence, but suggest this result may be better interpreted as $f_{\mathrm{NL}}^{\mathrm{local}} < 120$ at 84\% confidence, in light of the systematic concerns. General relativistic (GR) corrections are also expected to cause features in the clustering of galaxies at the largest (super-horizon) scales. These effects are expected to be small compared to that of $f_{\mathrm{NL}}^{\mathrm{local}}$, as, e.g., \cite{Maart12} find that the effects of $f_{\mathrm{NL}}^{\mathrm{local}}$ on the power spectrum dominate GR corrections for $f_{\mathrm{NL}}^{\mathrm{local}} \gtrsim 10$.

We analyze the SDSS data release nine (DR9; \citealt{DR9}) Baryon Oscillation Spectroscopic Survey (BOSS; \citealt{Daw12bosso}) ``CMASS'' sample of galaxies. This sample has the largest effective volume \citep{Teg98} of any spectroscopic survey and comprises approximately 1/3 of the final (planned) BOSS CMASS sample. The clustering of this data set has been extensively studied (\citealt{alph,Nuza12,ReidRSD12,SanchezCos12,Toj12RSD}). In particular, the sample has been simulated with 600 mock galaxy catalogs \citep{Manera12} and the issues identified by \cite{imsys} have been addressed via the application of an un-biased weighting scheme \citep{Ross12}. We thus have the tools to robustly investigate the information content in the large scale clustering, even in light of systematic concerns. 

Despite having the largest effective volume of any spectroscopic study, the DR9 CMASS sample is smaller than existing photometric redshift samples of galaxies and quasars (existing quasar samples have roughly 20 times the volume). Given that $f_{\mathrm{NL}}^{\mathrm{local}}$ imparts a change in the expected clustering measurements that is most pronounced at large scales and is smoothly varying, we should not expect $f_{\mathrm{NL}}^{\mathrm{local}}$ constraints from the DR9 CMASS sample to be competitive even with those one could obtain with the SDSS data release eight \citep{DR8} photometric redshift sample created in \cite{imsys} (which has approximately 3 times the angular footprint of our data set and was trained using early BOSS redshifts), let alone a photometric quasar sample. However, we expect our study to have best-quantified the impact of systematics, given the analyses of \cite{Ross12}. 

We primarily use measurements of the spherically-averaged power spectrum, $P(k)$, to constrain $f_{\mathrm{NL}}^{\mathrm{local}}$, as the scales most affected by local non-Gaussianity are most isolated in $k$-space. In Appendix \ref{app:xi}, we test for consistency using the spherically averaged correlation function, $\xi(s)$. The main purpose of this study is to demonstrate how the ability to constrain $f_{\mathrm{NL}}^{\mathrm{local}}$ using the galaxy power spectrum depends on the treatment of systematic uncertainties related to stellar density. We describe the observed and simulated data we use in Section \ref{sec:data} and our analysis methods in Section \ref{sec:analysis}. Our results, and their dependence on our treatment of potential systematics, are presented in Section \ref{sec:results}. In Section \ref{sec:discussion} we discuss our results in the context of current and future $f_{\mathrm{NL}}^{\mathrm{local}}$ measurements, and we summarize our conclusions in Section \ref{sec:con}.

Unless otherwise noted, we assume a flat cosmology with $\Omega_{m} = 0.285, \Omega_{b}=0.0459, h=0.70, n_s=0.96$, and $\sigma_8=0.8$, as are approximately the best-fit values found by \cite{SanchezCos12} when fitting the full shape of $\xi(s)$.

\section{Data}
\label{sec:data}
The SDSS-III Baryon Oscillation Spectroscopic Survey (\citealt{Eisenstein11}) obtains targets using SDSS imaging data. In combination, the SDSS-I, SDSS-II, and SDSS-III surveys obtained wide-field CCD photometry (\citealt{C,Gunn06}) in five passbands ($u,g,r,i,z$; \citealt{F}), amassing a total footprint of 14,555 deg$^2$, internally calibrated using the `uber-calibration' process described in \cite{Pad08}, and with a 50\% completeness limit of point sources at $r = 22.5$ (\citealt{DR8}). From this imaging data, BOSS has targeted 1.5 million massive galaxies, 150,000 quasars, and over 75,000 ancillary targets for spectroscopic observation over an area of 10,000 deg$^2$ \citep{Daw12bosso}. BOSS observations began in fall 2009, and the last spectra of targeted galaxies will be acquired in 2014. The BOSS spectrographs (R = 1300-3000) are fed by 1000 optical fibres in a single pointing, each with a 2$^{\prime\prime}$ aperture \citep{Smee12}. Each observation is performed in a series of 15-minute exposures and integrated until a fiducial minimum signal-to-noise ratio, chosen to ensure a high redshift success rate, is reached. Redshifts are determined as described in \cite{Bolton12}.

We use data from the SDSS-III DR9 BOSS CMASS sample of galaxies, as defined by \cite{Eisenstein11}. We use the same DR9 CMASS sample and treat it in the exact same way as in \cite{alph} and \cite{Ross12}. This sample has 264,283 galaxies spread over an effective area of 3275 deg$^2$, 2584 deg$^2$ of which is in the North Galactic Cap.  

We use the 600 mock DR9 CMASS catalogs (henceforth ``mocks'') generated by \cite{Manera12} to generate covariance matrices for clustering estimators. \cite{Manera12} used the initial conditions of a flat cosmology defined by $\Omega_{m}=0.274, \Omega_{b}h^2=0.0224, h=0.70, n_s=0.95$, and $\sigma_8=0.8$ (same as the fiducial cosmologies assumed in \citealt{White11BEDR} and \citealt{alph}) and generated dark matter halo fields at redshift 0.55. These simulations were produced using a 2nd-order Lagrangian Perturbation Theory (2LPT) approach inspired by the Perturbation Theory Halos (PTHalos; \citealt{PTHalos}) model. Galaxies were placed in halos using the halo occupation distribution determined from measurements of the correlation function of CMASS galaxies and the parameterization of \cite{ZCZ}. The DR9 angular footprint was then applied and galaxies were sampled along the radial direction such that the mean $n(z)$ matched the CMASS $n(z)$, thereby providing 600 catalogs simulating the observed DR9 CMASS sample. See \cite{Manera12} for further details. The mocks have $f_{\mathrm{NL}}^{\mathrm{local}} = 0$ and we describe how we  
account for this in our covariance matrices in Section \ref{sec:cov}.

\section{Analysis techniques}
\label{sec:analysis}
\subsection{Physical Model}
As shown in, e.g., \cite{Materrese00}, primordial non-Gaussianity alters the expected mass function of dark matter halos, with positive $f_{\mathrm{NL}}^{\mathrm{local}}$ yielding more high-mass dark matter halos. As outlined in both \cite{Dalal08} and \cite{Matt08}, this causes a scale dependent bias in the power spectrum of dark matter halos. These studies have shown that, expressing the Gaussian halo bias as $b$, the additive bias is given by
\begin{equation}
B_{NL}(k) = (b-1)f_{\mathrm{NL}}^{\mathrm{local}}A(k)
\label{eq:Bfnl}
\end{equation}
where\footnote{ Note, a factor of $h^2$ is required if one uses $k$ in units $h$Mpc$^{-1}$.}
\begin{equation}
A(k) = \frac{3\delta_c(z)\Omega_m}{k^2T(k)}\left(\frac{H_0}{c}\right)^2,
\label{eq:Ak}
\end{equation}
and $\delta_c(z) = 1.686/D(z)$ is the critical spherical over-density for dark matter halos to collapse at redshift $z$, $D(z)$ is the linear growth factor (normalized to equal 1 at $z=0$), and $T(k)$ is the transfer function. We obtain the $z=0$ linear power-spectrum, $P_M(k)$, and transfer function using the software package CAMB\footnote{see the website camb.info} \citep{camb} and our fiducial cosmology. 

The total bias is given by 
\begin{equation}
b_{tot} = b + B_{NL}(k).
\end{equation}
 Thus, the galaxy power spectrum, $P_g(k)$ can be expressed given the matter power spectrum, $P_M(k)$, as 
\begin{equation}
\begin{array}{l}
P_g(k) = b^2P_{M}(k) +2bB_{NL}(k)P_{M}(k)\\
+ B^2_{NL}(k)P_{M}(k).
\end{array}
\label{eq:Pexp}
\end{equation}
We will measure spherically averaged clustering in redshift space. Assuming linear theory, the redshift space, galaxy power spectrum, $P^o_g(k)$ is determined from the real space galaxy power spectrum \citep{Kaiser}
\begin{equation}
P^o_g(k) = P_{g}(k)+2/3fb_{tot}P_{M}(k)+1/5f^2P_{M}(k),
\label{eq:Pgo}
\end{equation}
where $f \equiv {\rm d ln}D/{\rm d ln}a$. The redshift zero result is then evolved via $P^o_g(k,z) = D^2(z)P^o_g(k,0)$. We restrict our analysis to $k < 0.05 h$Mpc$^{-1}$, where our linear model is expected to be a valid approximation, as is justified via comparison to the results of mock DR9 CMASS samples described in Section \ref{sec:pmeas}. 

Various studies (e.g., \citealt{Desj09,Wagner12,Scocc12}) have shown that non-zero $f_{\mathrm{NL}}^{\mathrm{local}}$ produces corrections to the scale-dependent and -independent halo bias that are more complicated functions of halo mass (and age) than the simple model we have presented above. Further, \cite{Desj11} and \cite{Roth12} show that higher-order contributions to the (non-Gaussian) primordial potential may bias determinations of $f_{\mathrm{NL}}^{\mathrm{local}}$. These concerns suggest that measurements of $f_{\mathrm{NL}}^{\mathrm{local}}$ using the techniques we describe above may be biased and that any precise determination of $f_{\mathrm{NL}}^{\mathrm{local}}$ must consider these effects. However, in all of the models, detection of non-zero $f_{\mathrm{NL}}^{\mathrm{local}}$ should occur only when there is local non-Gaussianity in the primordial potential. The values we determine should be viewed as effective $f_{\mathrm{NL}}^{\mathrm{local}}$ values, but we expect the relative importance of systematic biases imparted by observational effects to be independent of these model concerns.

\subsection{Measuring the Power Spectrum}
\label{sec:pmeas}
We measure the power spectrum, $P_m(k)$, using the standard Fourier technique of \cite{FKP}, as described in \cite{ReidDR7} and \cite{alph}. In particular, we calculate the spherically-averaged power in $k$ bands of width $\Delta k = 0.004h$Mpc$^{-1}$ using a 2048$^3$ grid. For all measurements we present, the galaxies are weighted to account for missing close-pairs (the BOSS spectrographs cannot simultaneously observe objects separated by less than 62$^{\prime\prime}$) and redshift failures. We also apply weights to account for the systematic relationship between target galaxy density and stellar density, and compare to results when these weights are not applied. We label measurements made using the stellar density weights as $P_{m,star}$, and treat these as the fiducial measurements. We label those made without the stellar density weights as $P_{m,nw}$. Finally, we apply `FKP' (standing for Feldman-Kaiser-Peacock) weights to both galaxies and randoms using the prescription of \cite{FKP}, which amounts to a redshift-dependent weighting in our application. The total weights for each galaxy/random are summed at each grid point. The process of calculating weights is described in detail in \cite{Ross12}. 

In the resulting measurements, $k$ bands will be correlated as a result of the survey geometry. We measure this effect by determining the power spectrum of the survey geometry, as simulated using an unclustered random catalog. This exercise yields the spherically-averaged ``window power'', which we denote $P_{win}(k)$. To combine the windows of the NGC and SGC, we determine the volume-weighted average of the respective $P_{win}(k)$.

\subsection{Accounting for the Survey Window}
\begin{figure}
\includegraphics[width=84mm]{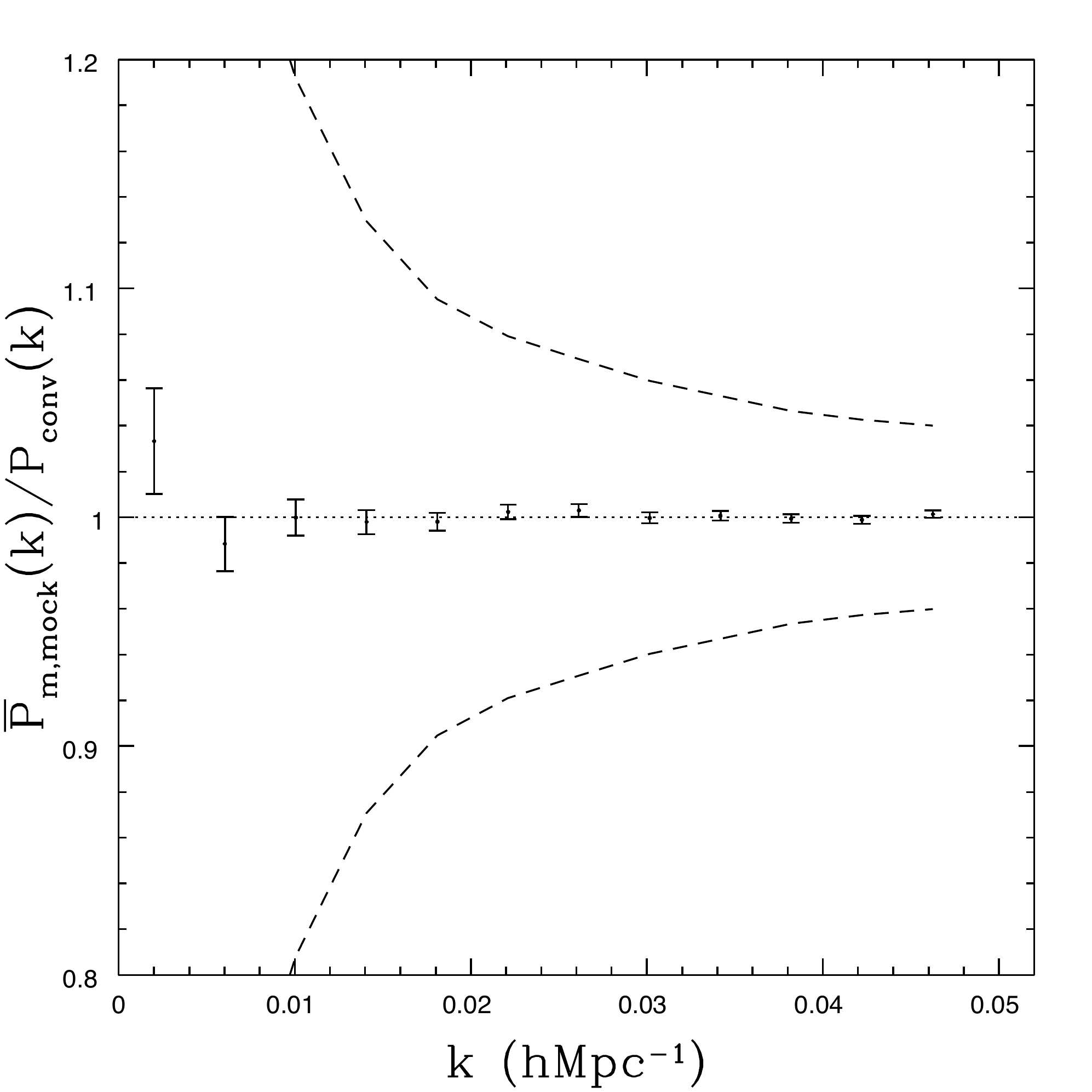}
  \caption{The mean power spectrum recovered from the mocks, $P_m(k)$, divided by the input power spectrum used to generate the mock galaxy catalogs that has been first translated to redshift space at $z=0.55$ assuming a real-space bias of 1.889 (the best-fit value) and then convolved with the window applied to the mock footprint. The error-bars are the 1$\sigma$ uncertainties on the mean of the 600 mocks and the dashed lines reflect the standard deviation of the 600 mocks.}
  \label{fig:pkconv}
\end{figure}

The measured $P_m(k)$ is a convolution between the window power and the true, underlying, power spectrum, $P_t(k)$. In practice, we measure $P_m(k)$ in discrete $k$-bands. Thus, we determine a ``window matrix'', $W[k_i][k_j]$, using the spherically-averaged $P_{win}$ and the process outlined in Appendix \ref{app:window}. Our expectation for the measured $P_{conv}(k)$ is given by
\begin{equation}
P_{conv}(k_i) = \sum_j W[k_i][k_j]P^o_{g,t}(k_j) - P_oP_{win}(k_i)
\label{eq:pkm}
\end{equation}
where
\begin{equation}
P_o = \sum_jW[0][k_j]P^o_{g,t}(k_j)_t/P_{win}(0).
\label{eq:p0}
\end{equation}
The second term, which we will here after refer to as the ``window subtraction'', in Eq. \ref{eq:pkm} is necessary because the measured $P_m(k)$ is zero at $k=0$ by definition\footnote{This is analogous to the integral constraint on the correlation function.}. Inspection of Eqs. \ref{eq:pkm} and \ref{eq:p0} reveals that the normalization given by $P_o$ yields $P_{conv}(0) = 0$, and thus $P_{conv}(0) = P_m(0) = 0$ by construction. We measure $P_{win}$ and calculate $P^o_g(k)$ in bins $\Delta k = 0.0005h$Mpc$^{-1}$, yielding sufficient resolution to calculate $P_{conv}(k)$ in (the measured) bin width $\Delta k = 0.004h$Mpc$^{-1}$.  

\begin{figure}
\includegraphics[width=84mm]{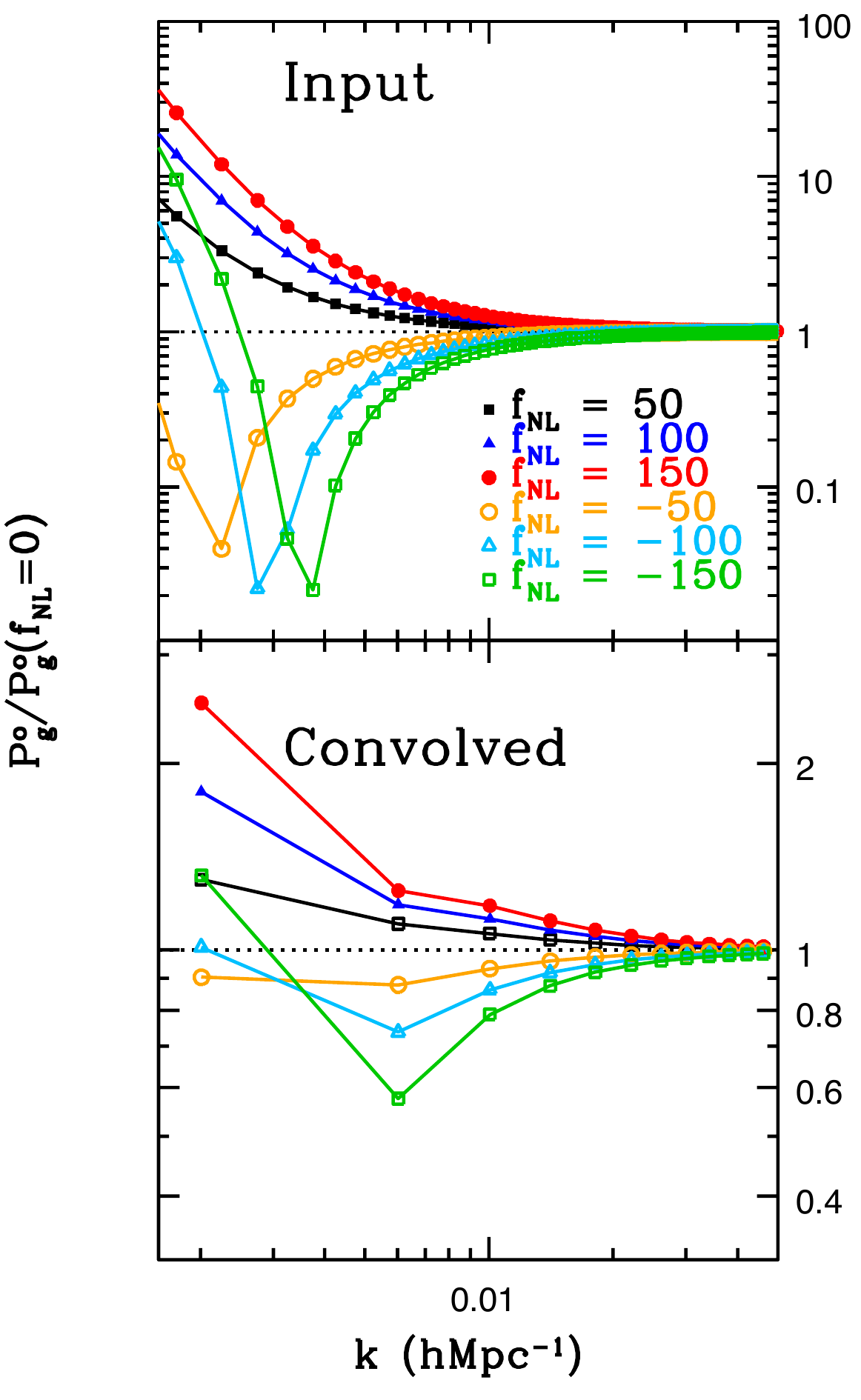}
  \caption{Top panel: The input redshift space power spectrum at $z=0.55$ assuming a real-space bias of 1.889 for the six labeled $f_{\mathrm{NL}}^{\mathrm{local}}$ values, divided by $P^o_g(k)$ for $f_{\mathrm{NL}}^{\mathrm{local}} = 0$. Bottom panel: The same information as the top panel, except the models have now been convolved with the window function and the range of the $P^o_g(k)/P^o_g(k,f_{\mathrm{NL}}^{\mathrm{local}}=0)$ axis has been decreased by more than an order of magnitude.}
  \label{fig:pkthconv}
\end{figure}

We test our determinations of $P_{win}$ and $W[k_i][k_j]$ using $P_m(k)$ measured from 600 mock galaxy catalogs generated by \cite{Manera12}. We should be able to recover the mean of the $P_m(k)$ measured for the mocks by convolving the input $P(k)$ used to create the mocks and the $P_{win}$ measured from the window applied in \cite{Manera12}. The ratio between the mean $P_m(k)$ recovered from the mocks and the convolved input power spectrum, $P_{conv}(k)$ (translated to redshift space at $z=0.55$ and assuming the cosmology used to generate the mocks and a real-space bias of 1.891), is displayed in Fig. \ref{fig:pkconv}. The error-bars reflect the uncertainty in the mean, and the dashed lines display the standard deviation (which is the expected statistical uncertainty for our CMASS measurements). We determine a best-fit bias of 1.891 by scaling the full covariance matrix (i.e., we divide each element by 600). Only for the bin centred at $k=0.002h$Mpc$^{-1}$ does the result differ by more than 1$\sigma$ from this best-fit theoretical expectation. However, the correlation between $k$-bins is significant and the minimum $\chi^2$ is 21.1 (11 degrees of freedom and 12 $k$-bins). The $\chi^2$ value is small enough to suggest our measurements of $f_{\mathrm{NL}}^{\mathrm{local}}$ should be unbiased. Indeed, we find $f_{\mathrm{NL}}^{\mathrm{local}} = $  0.6 $\pm$ 2.5 when we fit the average of the mock $P(k)$. Our treatment of the window therefore introduces no measurable bias on the recovered $f_{\mathrm{NL}}^{\mathrm{local}}$. This uncertainty of 2.5 corresponds to the mean of the 600 realizations of the DR9 CMASS sample; we therefore should expect a 1$\sigma$ uncertainty on $f_{\mathrm{NL}}^{\mathrm{local}}$ close to  60 ($\simeq$2.5$\times \sqrt{600}$) for the actual DR9 CMASS sample.

The window, and the associated window subtraction term in Eq. \ref{eq:p0}, imply that the expected measurement of the power spectrum will not have the same shape as the true underlying power spectrum. This effect is illustrated by Fig. \ref{fig:pkthconv}. In the top panel, a series of input $P^o_g(k)$ (evaluated at the centre of the $\Delta k =0.0005 h$Mpc$^{-1}$ bins) are plotted for -150 $\leq f_{\mathrm{NL}}^{\mathrm{local}} \leq$ 150, divided by the input $P^o_g(k)$ with $f_{\mathrm{NL}}^{\mathrm{local}} = 0$. At the lowest $k$, negative $f_{\mathrm{NL}}^{\mathrm{local}}$ actually causes an enhancement of the power spectrum; this is due to the fact that the term proportional to the square of $f_{\mathrm{NL}}^{\mathrm{local}}$ dominates and thus the square of the total bias becomes larger than in the case of $f_{\mathrm{NL}}^{\mathrm{local}}=0$. The bottom panel displays the same information as the top panel, after $P^o_g(k)$ has been convolved with the window (now for $\Delta k =0.002$ bins). At the low $k$ bins, the effect of $f_{\mathrm{NL}}^{\mathrm{local}}$ is decreased by more than an order of magnitude when compared to the input $P^o_g(k)$. At these values of $k$, the measurement depends strongly on the underlying $P^o_g(k)$ at larger $k$ and $P_{win}$, and their ability to constrain $f_{\mathrm{NL}}^{\mathrm{local}}$ is much weaker than one naively expects.

\subsection{Modelling the Systematic Contribution}
\label{sec:cov}
We compare the measured power spectrum, $P_m(k)$, of CMASS galaxies to model $P_{mod}(k,f_{\mathrm{NL}}^{\mathrm{local}},S)$, which combine the theoretical galaxy power spectrum convolved with the CMASS window function and a term $S$ (described below) to model the contribution of systematics to the measured power. \cite{Ross12} found that a weight for stellar density had a large effect on the measured low-$k$ CMASS power spectrum. We assume that the correction for potential systematics always has the same form, but the magnitude of the correction can vary. This assumption is justified by the fact that we found this to be true when testing different weighting schemes in \cite{Ross12}. Our total model is 
\begin{equation}
P_{mod}(k,f_{\mathrm{NL}}^{\mathrm{local}},S) = P_{conv}(k,f_{\mathrm{NL}}^{\mathrm{local}})+S[P_{m,nw}(k)-P_{m,star}(k)],
\label{eq:pmod}
\end{equation}
where $P_{m, star}(k)$ denotes the measurement made using the stellar density weight, and $P_{m,nw}(k)$ denotes the measurement determined when the stellar density weights are not applied. Treating $P_{m,star}$ as the fiducial measurement, this approach fixes the systematic correction in the observed power spectrum and uses the model to account for the possibility that our measurement has a remaining systematic bias by scaling the amplitude of the fiducial correction. Specifically, $S=0$ represents the fiducial case assuming we have properly removed any systematic bias from our measurement and positive/negative $S$ accounts for the possibility a systematic bias remains/has been incorrectly subtracted by our treatment of the data. 

We apply weights to account for the systematic effect of stars when we determine the power spectrum. These weights were determined via a linear fit to the relationship between the target galaxy density and the stellar density. We can estimate the uncertainty on $S$ by determining these weights for each of the 600 mocks (which have no intrinsic systematic relationship). This analysis was performed in \cite{Ross12}, where $\xi(s)$ was measured for each mock sample with weights determined and applied in each individual case. The weights did not bias the mean of the measurements or the variance (see their Fig. A2), but each individual measurement was given a nearly constant bias with standard deviation 1.5$\times10^{-4}$. This standard deviation on the bias is 10\% the size of difference between the unweighted, $\xi_{nw}$, and weighted CMASS $\xi(s)$. Therefore, we expect the statistical uncertainty on $S$ to be 0.1 for the DR9 CMASS sample, and we believe allowing any value of $S$ accounts for additional, unknown systematic effects. 

\subsection{Comparing Measurement to Model}
Previous DR9 studies (e.g., \citealt{alph,Ross12}) used measurements of the power spectra of 600 mock realizations of the CMASS sample to construct the covariance matrix. These  mock realizations are for $f_{\mathrm{NL}}^{\mathrm{local}} = 0$. For a constant number density, $\bar{n}$, the total variance is expected to scale as (see, e.g., \citealt{FKP})
\begin{equation}
\sigma^2_P \propto (P+1/\bar{n})^2. 
\label{eq:psig}
\end{equation}
The value of $f_{\mathrm{NL}}^{\mathrm{local}}$ significantly alters the expected power spectrum at low $k$, and thus we expect the amplitude of the covariance to depend significantly on the true value of $f_{\mathrm{NL}}^{\mathrm{local}}$. The number density for the DR9 CMASS sample is $\sim3\times10^{-4} h^{3}$Mpc$^{-3}$ and its measured power spectrum is greater than 2.5$\times10^4 h^{-3}$Mpc$^3$ across all of the scales we consider. Thus, we expect cosmic variance to dominate the uncertainty at these scales by a factor of at least 8 (the ratio of $P$ to $1/\bar{n}$).  Therefore, the expected uncertainty on the power spectrum of can be approximated by $\sigma_P \propto P$ and thus the expected uncertainty on $\sigma_p/P$ can be approximated as independent of the theoretical model.

The uncertainty on ${\rm ln}[P]$ is $\frac{\sigma_P}{P}$. Therefore, and based on the arguments above, we compare ${\rm ln}[P_m]$ to ${\rm ln}[P_{mod}]$ and thus when testing models against our measurements, we minimize the $\chi^2$ given by
\begin{equation}
\chi^2 = \sum_{i,j} {\rm ln}\left[\frac{P_m(k_i)}{P_{mod}(k_i)}\right] C^{-1}_p[k_i][k_j] {\rm ln}\left[\frac{P_m(k_j)}{P_{mod}(k_j)}\right],
\end{equation}
where $C^{-1}_p$ is the covariance matrix of ${\rm ln}[P_m]$. $C^{-1}_p$ is determined from the mocks via
\begin{equation}
C_p[k_i][k_j] =
\frac{1}{599}\sum_{n=1}^{600}{\rm ln}\left[\frac{P_m^n(k_i)}{\bar{P}_m(k_i)}\right]{\rm ln}\left[\frac{P_m^n(k_j)}{\bar{P}_m(k_j)}\right],
\end{equation}
where $P_m^n$ represents the measured power spectrum of mock $n$ and $\bar{P}_m$ represents the mean across all mocks.

For the linear theory power spectrum, the correlation between different $k$-bins is the effect of the survey window alone. Thus, in the cosmic variance limit (in which the diagonal elements are constant), the covariance matrix of the logarithm of the power for $k < 0.05 h$Mpc$^{-1}$ depends primarily on the survey window. Thus, to the extent that we can ignore model dependent variations in the shot-noise term and non-linear effects on the covariance of ${\rm ln}[P]$  for $k < 0.05 h$Mpc$^{-1}$, we expect our treatment to be correct independent of $f_{\mathrm{NL}}^{\mathrm{local}}$, or any other cosmological model that changes the expected amplitude of the power spectrum on large scales\footnote{Our covariance matrix does include both shot-noise and non-linear effects, it simply will not account for the possibility that these effects scale {\it differently} with changes in, e.g., $f_{\mathrm{NL}}^{\mathrm{local}}$ than the power.}. Our choice to test ln$[P(k)]$ is further supported by the fact that, as described in Appendix \ref{app:mocktest}, we find that the skewness of the distribution of {\rm ln}$[P(k)]$ is significantly smaller than that of $P(k)$.

We are using the same $P_m(k)$ measurements as \cite{alph}, which assumed a flat geometry with $\Omega_m = 0.274$, but we adopt a flat geometry with $\Omega_m = 0.285$ as our fiducial cosmology. In order to account for the difference, we assume spherical symmetry and dilate the center of the $k$-bin by $D_v(\Omega_m=0.285)/D_v(\Omega_m=0.274) = 1.006$, where
\begin{equation}
D_v = \left[cz(1+z)^2D^2_AH^{-1}\right]^{1/3},
\end{equation}
where $D_A$ is the angular diameter distance and $H$ is Hubble parameter. Thus, when comparing the theoretical power spectrum to our measurements, we determine the input $P^o_g(k)$ in $k$-bins of width 5.03$\times 10^{-4}h$Mpc$^{-1}$ and the convolved $P_{conv}(k)$ in $k$-bins of width 4.024$\times 10^{-3}h$Mpc$^{-1}$ (whereas the $k$-bin sizes were 5$\times 10^{-4}h$Mpc$^{-1}$ and 4$\times 10^{-3}h$Mpc$^{-1}$ for the flat $\Omega_m = 0.274$ cosmology). Dilations are applied in the same manner when alternative values of $\Omega_m$ are tested.

\section{Results}
\label{sec:results}

\begin{table*}
\begin{minipage}{7in}
\caption{The recovered $f_{\mathrm{NL}}^{\mathrm{local}}$ constraints and quality of fit for the four treatments of the systematic contribution to the model that we consider, where C.I. stands for confidence interval, ML stands for maximum likelihood, and $P(f_{\mathrm{NL}}^{\mathrm{local}}<0)$ is the sum of the probability at $f_{\mathrm{NL}}^{\mathrm{local}}$ less than zero.}
\begin{tabular}{lllccclc}
\hline
\hline
Case  &  $f_{\mathrm{NL}}^{\mathrm{local}}$ 95\% C.I. &  $|f_{\mathrm{NL}}^{\mathrm{local}}|$ 95\% C.I. &  $f_{\mathrm{NL}}^{\mathrm{local}}$ ML & overall best-fit $f_{\mathrm{NL}}^{\mathrm{local}}$  & $\chi^2_{\rm best-fit}$/dof & $S_{\rm best-fit}$ & $P(f_{\mathrm{NL}}^{\mathrm{local}}<0)$ \\
\hline
i & (-27, +196) & (0, 184)  & 106 & 105 & 15.9/10 & $\equiv 0$ & 5.9\%\\
ii & (+32, +198) & (32, 198)& 123 & 123 & 32.4/10 & $\equiv -1$ & 0.5\%\\
iii & (-45, +195) & (0, 179)  & 102 & 105 & 15.6/9 & 0.08 & 9.0\%\\
iv & (-82, +178) & (0, 154) & 76 & -48 & 13.0/9 & 0.45 & 32\%\\

\hline
\label{tab:allpar}
\end{tabular}
\end{minipage}
\end{table*}

\begin{figure}
\includegraphics[width=84mm]{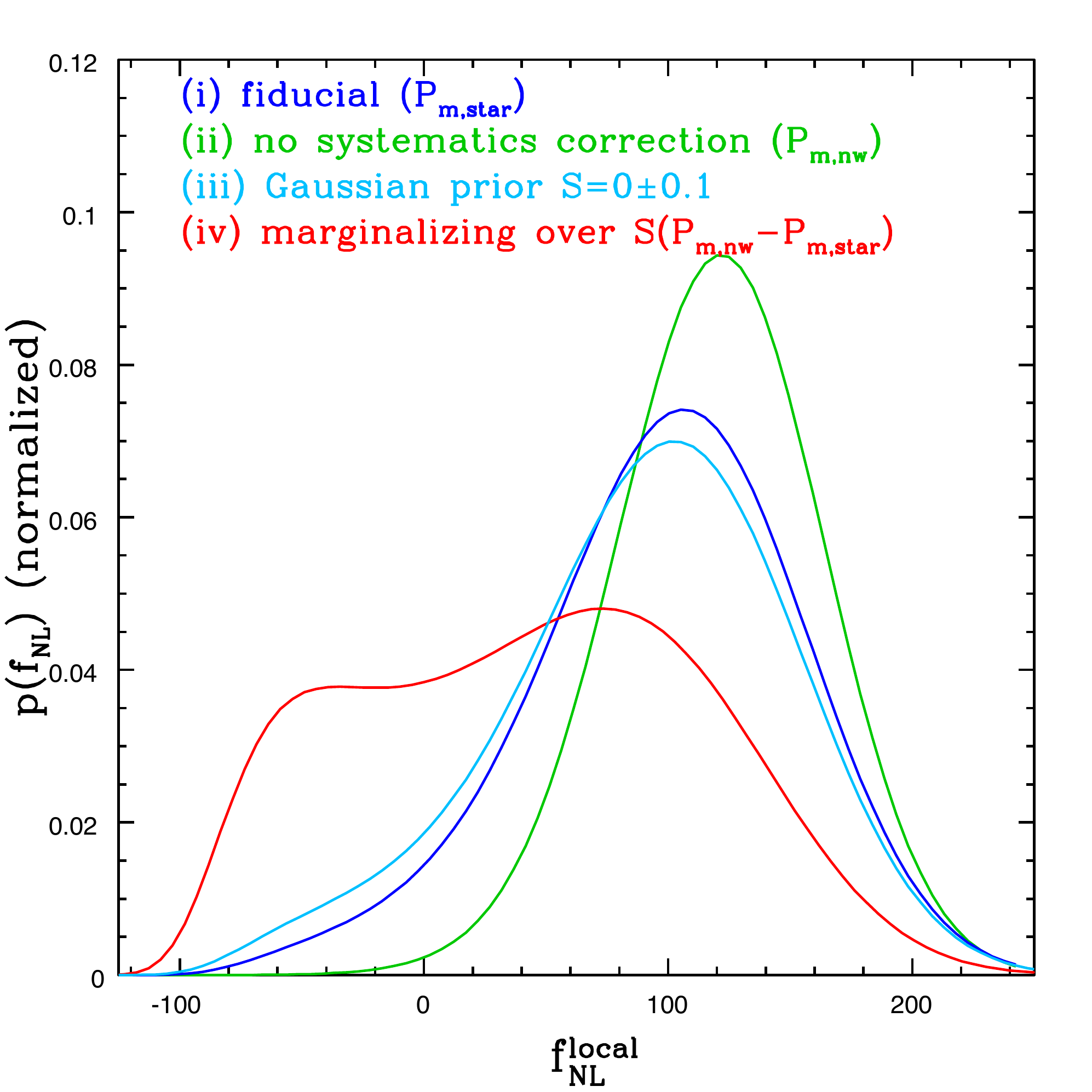}
  \caption{The normalized (so that they integrate to 1) probability distributions for the local non-Gaussianity parameter $f_{\mathrm{NL}}^{\mathrm{local}}$,  for our four treatment of systematics, applied to the DR9 CMASS sample. The blue curve shows the result using our fiducial treatment which uses the power spectrum, $P_{m,star}$ determined using the stellar density weights; the green curve shows the result when we use the power spectrum measurement, $P_{m,nw}$, that does not include the stellar density weights; the light blue curve shows the result when we marginalize over an additional term $S(P_{m,nw}-P_{m,star})$ in the model and allow it vary within a Gaussian prior of $0\pm0.1$; and the red curve shows the results when we marginalize over $S$ and allow it vary freely.}
  \label{fig:pfnl}
\end{figure}

\begin{figure}
\includegraphics[width=84mm]{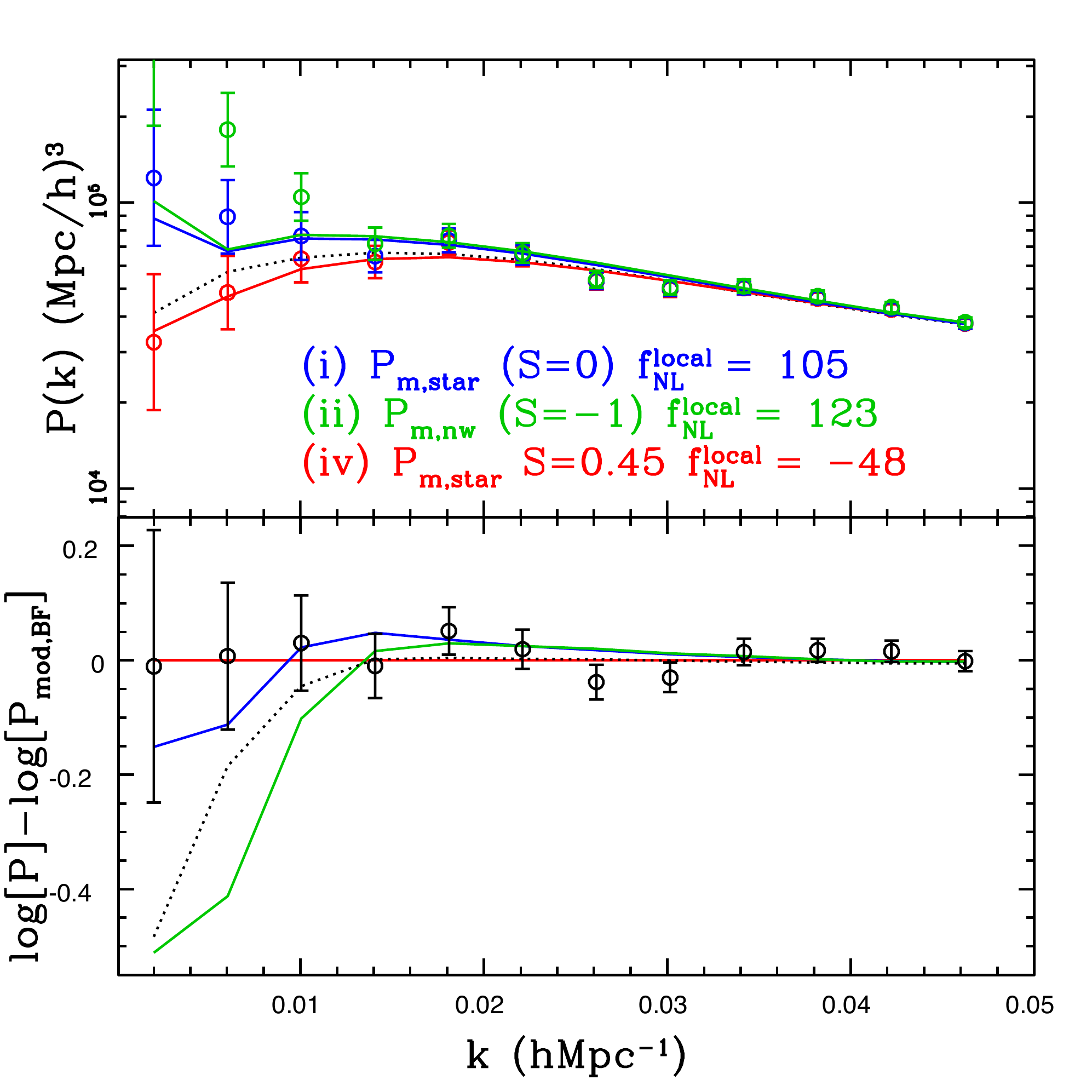}
  \caption{Top panel: The measured DR9 CMASS $P(k)$ for the labeled treatment of systematics (points with error-bars) and the associated best-fit model (solid lines). The model with $f_{\mathrm{NL}}^{\mathrm{local}}=0$ and $S=0$ is displayed with a dotted line. We subtract $S(P_{m,nw}-P_{m,star})$ from the measured power spectrum, rather than add it to the theoretical model; in this way the points show the measurement assuming the given systematic treatment is the correct one. Bottom panel: The difference between the logarithms of a given power spectrum and the overall best-fit model power spectrum (which is for $S=0.45$, $f_{\mathrm{NL}}^{\mathrm{local}} =$ {\bf -48}), where the black circles represent the measured power spectrum (using weights for stellar density, $P_{m,star}$) and the lines represent the same four models and use the same scheme as in the top panel.}
  \label{fig:pkfnl}
\end{figure}

We present results for four treatments of the systematic contribution to the model described by Eq. \ref{eq:pmod}: \begin{enumerate}  
\item $S=0$; this corresponds to our best estimate of the power spectrum, measured using the weights for stellar density, but we ignore systematic uncertainty in the application of these weights; 
\item $S=-1$; this corresponds to the case where no weights are applied for stellar density;
 \item $S$ is allowed to vary around 0 with a Gaussian prior of width 0.1 (this value is as motivated by the tests performed on mocks in \citealt{Ross12}) accounting for the uncertainty in application of the stellar density weight; and  
 \item Any value $S > -1$ is permitted. This final case assumes we have learned the $k$-dependence of systematic effects on our measurement, but allows for the possibility of an unknown systematic ($S>0$) or that the stellar density weights were an over-correction ($S<0$).
 \end{enumerate} 
 We marginalize over $S$ for cases (iii) and (iv) and $b$ in all cases. There is a small degeneracy between $\Omega_m$ and $f_{\mathrm{NL}}^{\mathrm{local}}$. We find that the best-fit value of $f_{\mathrm{NL}}^{\mathrm{local}}$ shifts by 6 comparing case (i) with $\Omega_m = 0.26$ and $\Omega_m = 0.3$. This change is approximately 10\% of the expected statistical uncertainty, and is thus small enough that we expect the conclusions of this study are robust to changes in the background $\Lambda$CDM cosmology (flat $\Lambda$CDM with $\Omega_{m} = 0.285, \Omega_{b}=0.0459, h=0.70, n_s=0.96$, and $\sigma_8=0.8$).

Fig. \ref{fig:pfnl} displays the recovered probability distribution functions (PDFs) when we fit our power-spectrum measurements using the four treatments of the systematics we consider, and Table \ref{tab:allpar} lists the relevant attributes in each case. Only for case (ii), where we use the measurement without the weights for stellar density, (displayed in green) does the probability distribution appear Gaussian. For this case, $f_{\mathrm{NL}}^{\mathrm{local}} < 0$ is allowed at 0.5\% and the PDF peaks at $f_{\mathrm{NL}}^{\mathrm{local}} =$ 123. In the absence of any systematic correction, we would have appeared to detect non-zero $f_{\mathrm{NL}}^{\mathrm{local}}$. However, the value of the $\chi^2$ at the minimum, $\chi^2_{\rm best-fit}= 32.4$ (10 degrees of freedom), for case (ii) indicates a problem as only 0.03\% of samples consistent with our model would yield as large a $\chi^2$. This result, in and of itself, suggests the analysis performed in case (ii) has a systematic in the treatment of the data/modelling.


Including the weights for stellar density, case (i), reduces the $\chi^2$ minimum to 15.9 (10 degrees of freedom), and we expect a $\chi^2$ value at least as large for 10\% of consistent samples. The $\Delta \chi^2_{\rm best-fit} = 16.5$ between cases (ii) and (i) shows that the stellar density weights are strongly preferred. However, the maximum likelihood value of $f_{\mathrm{NL}}^{\mathrm{local}}$ decreases only from 123 to 106. One can see in the top panel of Fig. \ref{fig:pkfnl} that the convolved models using these two $f_{\mathrm{NL}}^{\mathrm{local}}$ values (plotted so that the systematic term is applied to the measurement, not the model) appear qualitatively similar. The measurements at scales $k > 0.02h$Mpc$^{-1}$ (where the stellar density weights are not important) do not accommodate a significantly larger $f_{\mathrm{NL}}^{\mathrm{local}}$; it is clearly the tension between the power spectrum measurements in $k$ bins greater and less than $k = 0.02h$Mpc$^{-1}$ that yields such a large $\chi^2$ in case (ii).

The comparison between cases (i) and (iii) illustrates the importance of including the uncertainty on the systematic correction. In both cases, we use the measured power spectrum that includes the weights for stellar density. In case (iii), we are essentially adding a systematic uncertainty, by allowing $S$ to vary around 0 with a Gaussian prior of width 0.1 (which is the uncertainty we determine from DR9 mocks), while for case (i) $S$ is simply set to zero, meaning we have applied a systematic correction but allowed no uncertainty due to the correction. Comparing the PDF of case (iii) to that of case (i), we find that the PDF widens such that the 95\% confidence interval increases by 17 (representing an 8\% increase), its peak shifts by $\Delta f_{\mathrm{NL}}^{\mathrm{local}} =$ 4, and the probability that $f_{\mathrm{NL}}^{\mathrm{local}} < 0$ increases from 5.9\% to 9.0\%. While these changes are not dramatic, they are significant and suggest (assuming the uncertainties add in quadrature so that $\sigma^2_{sys} = \sigma^2_{tot}-\sigma^2_{stat}$) that the 1$\sigma$ systematic uncertainty on $f_{\mathrm{NL}}^{\mathrm{local}}$ for the DR9 CMASS sample is $\sim$ 22. 

In case (iv), we allow any value of $S$. As shown in Fig. \ref{fig:pfnl}, 32\% of the PDF (displayed in red) is now at $f_{\mathrm{NL}}^{\mathrm{local}} < 0$. Further, there are two peaks; the most likely value is at $f_{\mathrm{NL}}^{\mathrm{local}} =$ 76, but there is a second peak at $f_{\mathrm{NL}}^{\mathrm{local}} =$ -31 with an amplitude that is 79\% that of the overall peak. As described in Appendix \ref{app:mocktest}, when case (iv) was tested on the 600 mock realizations, multiple peaks are found in 24\% of the PDFs (and in only 9\% of the PDFs for the other cases). The minimum $\chi^2$ for case (iv), 13.0 (9 degrees of freedom), is the overall minimum across all cases. The minimum occurs when $S = 0.45$ and $f_{\mathrm{NL}}^{\mathrm{local}} =$ -48. Inspection of Fig. \ref{fig:pkfnl} reveals that the negative $f_{\mathrm{NL}}^{\mathrm{local}}$ value allows a better fit to the two measurements over $0.024 < k < 0.03h$Mpc$^{-1}$ and the 45\% stronger correction allows the measurements at large-scales to be in good agreement. The $\chi^2$ is reduced by 2.9 with the addition of one extra parameter (comparing to case i), and we would expect 16\% of consistent samples to have a larger $\chi^2$.

\section{Discussion}
\label{sec:discussion}
\subsection{The Meaning of Our Results}
The exact treatment of systematics causes large differences in the allowed range of $f_{\mathrm{NL}}^{\mathrm{local}}$, as the width of the 95\% confidence interval increases from 167 to 261 from the least (case ii) to most (case iv) conservative treatment of systematics. The results of \cite{Ross12} suggest a 10\% uncertainty on the magnitude of the systematic correction applied for stellar density, and our case (iii) takes this effect into account, representing our treatment of systematics including everything we {\it know}. The 95\% confidence interval is -45 $< f_{\mathrm{NL}}^{\mathrm{local}} <$ 195. 

A more pessimistic approach assumes that we know only the way that the systematic correction scales with $k$, but not its amplitude. This approach is treated in case (iv) and allows for the presence of an {\it unknown} systematic. The 95\% confidence interval for this situation is -82$< f_{\mathrm{NL}}^{\mathrm{local}} <$178. The width of the 95\% confidence interval increases by only 9\% with respect to case iii. This result implies that more extreme values of $f_{\mathrm{NL}}^{\mathrm{local}}$ are rejected by the $P(k)$ measurements at the scales relatively unaffected by the systematic corrections. The tests of our methods on the mocks (in Appendix \ref{app:mocktest}) suggest that, for the DR9 CMASS sample, the confidence interval of the absolute value of $f_{\mathrm{NL}}^{\mathrm{local}}$ is most robust when less than 90\% of the total probability is contained on one side of zero (i.e., $10\% < P(f_{\mathrm{NL}}^{\mathrm{local}}<0) < 90\%$). We find $|f_{\mathrm{NL}}^{\mathrm{local}}| <$ 154 at 95\% confidence for case (iv).


The minimum $\chi^2$ decreases by 2.9 when we allow $S$ to freely vary (the minimum is $S=0.45$) compared to when we set $S=0$, suggesting that there there may be a systematic in the treatment of the data that is undiscovered. We do not find that any smoothly varying model for $P^o_g$ can produce a similar decrease in the $\chi^2$. For example, if the index of the scale-dependent bias is allowed to vary freely, we find the minimum $\chi^2$ decreases by only 1.0 (when the scaling is $\Delta b \propto k^{-1.8}$). We have taken a closer look at all of the potential systematics studied in \cite{Ross12} and found no further systematic effects on the measured clustering. We have also investigated the evolution of the clustering amplitude in the sample (a strongly evolving amplitude would preferentially increase the clustering at large scales) and found the variation to be less than 10\%, implying it has a negligible impact on our measurements. We suggest that this issue be revisited in future BOSS data releases, where the preference for a stronger correction should become more significant if there is an undiscovered systematic and there will be more data with which to test against potential systematic effects.

Other studies have obtained much tighter constraints on $f_{\mathrm{NL}}^{\mathrm{local}}$ using measurements of galaxy/quasar clustering: \cite{Slosar08} found $-29 < f_{\mathrm{NL}}^{\mathrm{local}} < 70$ and \cite{Xia11} found $5 < f_{\mathrm{NL}}^{\mathrm{local}} <  84$ (both at 95\% confidence). Both studies include SDSS photometric luminous red galaxy (LRG) samples that have a similar redshift range to that of our SDSS DR9 CMASS sample (and significant overlap in angular footprints). Using only the SDSS photometric LRG data sample defined by \cite{Pad07}, \cite{Slosar08} found $-268 < f_{\mathrm{NL}}^{\mathrm{local}} < 164$ while \cite{Xia11} measured $-81 < f_{\mathrm{NL}}^{\mathrm{local}} < 351$ using the LRG sample created in \cite{thomas10}. These 95\% confidence intervals are approximately twice as wide as we determine, but it is doubtful that substituting our own results for them would yield a significant change in any analysis that combines multiple samples, as most of the constraining power comes from the high redshift quasars. The final BOSS sample will be approximately three times as large as the DR9 sample and thus nearly halve (1/$\sqrt{3}$) the statistical uncertainty on $f_{\mathrm{NL}}^{\mathrm{local}}$ that we find.

\subsection{Lessons for the Future}

On its own, the BOSS DR9 CMASS sample does not allow significant improvement on existing $f_{\mathrm{NL}}^{\mathrm{local}}$ constraints. However, future studies can emulate our treatment of systematics uncertainties in order to obtain robust results. Here, we find that simply accounting for the 10\% uncertainty (implied by \citealt{Ross12}) in the {\it correction} we apply for stellar density increases the width of the 95\% confidence interval we determine for $f_{\mathrm{NL}}^{\mathrm{local}}$ by 17, implying $\sigma_{sys} =$ 22. It is not clear how the systematic uncertainty will scale in the presence of either more BOSS data or other data sets, e.g., this will depend on the exact distribution of stars across the angular footprint. It is therefore possible that the systematic uncertainty will become, relatively, more (or less) important for future CMASS data samples or alternative large-scale structure probes. We recommend future studies subject their data samples to the level scrutiny presented in \cite{imsys,Ross12}. We expect any systematic dependency on a Galactic foreground (e.g., stars, Galactic extinction, synchrotron emission) to impart a systematic bias and uncertainty onto $f_{\mathrm{NL}}^{\mathrm{local}}$ measurements; we caution that all future studies of large-scale clustering should test against this possibility in order to account for potential systematic errors.

Concerns regarding systematics are especially relevant to the photometric quasar data that, to date, represent the large-scale structure data sample that yields the best $f_{\mathrm{NL}}^{\mathrm{local}}$ constraints, as \cite{Xia11} find a 1$\sigma$ uncertainty on $f_{\mathrm{NL}}^{\mathrm{local}}$ of 26 when using only a SDSS data release six (DR6; \citealt{DR6}) photometric quasar sample. Photometric quasar samples are ideal for $f_{\mathrm{NL}}^{\mathrm{local}}$ studies, as the quasars occupy a large volume (decreasing the statistical uncertainty on large-scale measurements), have a large bias (increasing the relative effect of $f_{\mathrm{NL}}^{\mathrm{local}}$), and are at high redshift (and therefore have a smaller growth factor, which also increases the relative effect of $f_{\mathrm{NL}}^{\mathrm{local}}$). They are also known to be subject to stellar contamination: \cite{Myers07} found a stellar contamination of $4.4^{+1.9}_{-4.4}$\%, for a photometric SDSS data release four \citep{DR4} quasar sample constructed using the same methodology \citep{Richards04} as the quasar samples used by both \cite{Slosar08} and \cite{Xia11}. 

The 1$\sigma$ bounds, using only quasars, were found to be $8^{+26}_{-37}$ in \cite{Slosar08} (for their fiducial QSO case using SDSS data release three (DR3; \citealt{DR3}) quasars) and $62\pm26$ in \cite{Xia11} (using SDSS DR6 quasars). While the $f_{\mathrm{NL}}^{\mathrm{local}}$ constraints are impressive, the two results are discrepant at  $\sim1.5\sigma$ (while using much of the same data).  \cite{Slosar08} rejected quasars with photometric redshifts $z < 1.45$, after finding the cross correlation of this sample with stars had a non-zero amplitude. \cite{Xia11} made no such cut on their quasar sample since they found no evidence of different contamination when splitting the data at $z = 1.45$, and they suggested that the difference was due to better calibration of the DR6 data compared to the DR3 sample. The degree to which stellar contamination could explain the tension between the two results is studied further in Giannantonio et al. (in prep.), where methods similar to our own are used to fully account for the systematic uncertainty of the effect of Galactic foregrounds on the measured clustering of quasars (and other tracers) and to obtain robust constraints on $f_{\mathrm{NL}}^{\mathrm{local}}$.

\section{Conclusions}
\label{sec:con}
\noindent$\bullet$ We have described a method that quantifies both the systematic bias imparted by Galactic foregrounds on clustering measurements {\it and} its uncertainty, thus allowing one to obtain un-biased $f_{\mathrm{NL}}^{\mathrm{local}}$ constraints with realistic error estimates.  

\noindent$\bullet$ We find no reliable evidence for non-zero $f_{\mathrm{NL}}^{\mathrm{local}}$.

\noindent$\bullet$ The data show an extreme preference for our fiducial systematic correction (the application of weights for stellar density when calculating the power spectrum, case [i]) compared to the no systematics correction case (ii): the difference in the minimum $\chi^2$ is 16.5, when fitting for $f_{\mathrm{NL}}^{\mathrm{local}}$ in both cases.

\noindent$\bullet$ We have shown that the systematic effect of stars on the DR9 CMASS galaxy density field significantly affects the $f_{\mathrm{NL}}^{\mathrm{local}}$ constraints, but that the systematic bias has a different scale dependence than the (convolved) scale dependence of the effect of $f_{\mathrm{NL}}^{\mathrm{local}}$. Thus, each effect is detectable in the data, and $f_{\mathrm{NL}}^{\mathrm{local}}$ constraints can be obtained even when allowing any amplitude of the systematic correction (case iv), resulting in a 17\% increase in the width of the 95\% confidence interval.

\noindent$\bullet$ We find that the data exhibit a marginal preference for a stronger systematic correction than provided by our fiducial weights for stellar density, as the minimum $\chi^2$ decreases by 2.9 when the correction for stellar contamination is 45\% stronger. We find no physical model that produces a similar change in the minimum $\chi^2$, but also find no additional systematic effect that can explain the need for such a systematic correction.

We encourage all future studies to incorporate systematic uncertainties in a manner similar to that presented here in order to obtain $f_{\mathrm{NL}}^{\mathrm{local}}$ constraints that are robust to systematic uncertainties related to the treatment of the data.

\section*{Acknowledgements}
We thank the anonymous referee for comments that helped improve this paper.
AJR is grateful to the UK Science and Technology Facilities Council for financial support through the grant ST/I001204/1.
WJP is grateful for support from the the UK Science and Technology
Facilities Research Council, and the European
Research Council.

Part of the research described in this paper was carried out at the
JetPropulsion Laboratory, California Institute of Technology, under a
contract with the National Aeronautics and Space Administration.

Funding for SDSS-III has been provided by the Alfred P. Sloan
Foundation, the Participating Institutions, the National Science
Foundation, and the U.S. Department of Energy Office of Science.
The SDSS-III web site is http://www.sdss3.org/.

SDSS-III is managed by the Astrophysical Research Consortium for the
Participating Institutions of the SDSS-III Collaboration including the
University of Arizona,
the Brazilian Participation Group,
Brookhaven National Laboratory,
Cambridge University ,
Carnegie Mellon University,
Case Western University,
University of Florida,
Fermilab,
the French Participation Group,
the German Participation Group,
Harvard University,
UC Irvine,
Instituto de Astrofisica de Andalucia,
Instituto de Astrofisica de Canarias,
Institucio Catalana de Recerca y Estudis Avancat, Barcelona,
Instituto de Fisica Corpuscular,
the Michigan State/Notre Dame/JINA Participation Group,
Johns Hopkins University,
Korean Institute for Advanced Study,
Lawrence Berkeley National Laboratory,
Max Planck Institute for Astrophysics,
Max Planck Institute for Extraterrestrial Physics,
New Mexico State University,
New York University,
Ohio State University,
Pennsylvania State University,
University of Pittburgh
University of Portsmouth,
Princeton University,
UC Santa Cruz,
the Spanish Participation Group,
Texas Christian University,
Trieste Astrophysical Observatory
University of Tokyo/IPMU,
University of Utah,
Vanderbilt University,
University of Virginia,
University of Washington,
University of Wisconson
and Yale University.

\begin{appendix}

\section{Window Function Convolution}
\label{app:window}
Here we outline the process of determining $W[k_i][k_j]$, used in Eq. \ref{eq:pkm}, to obtain the model we compare to the measured power spectrum.

The ``spherically averaged power spectrum" within some $k$-volume $V_k$ can be modeled as the convolution between the true underlying power spectrum, and a window function, $W$  
\begin{equation}
P_m(k) = \frac{1}{V_k}\int_{V_k} d^3k^{\prime}\int\frac{d^3\epsilon}{(2\pi)^3}P_t(\vec{k}^{\prime}+\vec{\epsilon})W(\vec{\epsilon}).
\label{eq:ap1}
\end{equation}
We measure the power at particular $k$. Approximating the $k$-volume as a thin shell, $\frac{1}{V_k}\int_{V_k}d^3k^{\prime} = \int d\Omega_k\delta_{D}(k-|\vec{k}^{\prime}|)$, Eq. \ref{eq:ap1} can be rewritten
\begin{equation}
P_m(k) = \frac{1}{(2\pi)^3}\frac{1}{(4\pi)}\int d\Omega_k\int d\Omega_{\epsilon}\int dr_{\epsilon}P_t(\vec{k}+\vec{\epsilon})W(\vec{\epsilon}),
\end{equation}
where $r_{\epsilon} = |\vec{k}+\vec{\epsilon}|$. We change variables from $(\Omega_{\epsilon},\Omega_{k})$ to $(\Omega_{\epsilon}-\Omega_{k},\Omega_{\epsilon})$. Assuming isotropy, $P_t(\vec{k}+\vec{\epsilon})$ can be removed from the integral over $\Omega_{\epsilon}$ and we obtain
\begin{equation}
P_m(k) = \frac{1}{(2\pi)^3}\frac{1}{(4\pi)}\int dr_{\epsilon} \int d(\Omega_{\epsilon}-\Omega_k)P_t(\vec{k}+\vec{\epsilon})\int d\Omega_{\epsilon}W(\vec{\epsilon})
\end{equation}

We now define
\begin{equation}
W_s(\vec{\epsilon}) \equiv \frac{1}{4\pi}\int d\Omega_{\epsilon^{\prime}}W(\vec{\epsilon^{\prime}})\delta_D(r_{\epsilon^{\prime}}-r_{\epsilon}),
\end{equation}

\noindent which is independent of angle and obtain
\begin{equation}
P_m(k) = \frac{1}{(2\pi)^3}\int dr_{\epsilon}W_s(\vec{\epsilon}) \int d(\Omega_{\epsilon}-\Omega_k)P_t(\vec{k}+\vec{\epsilon})
\end{equation}
and thus
\begin{equation}
P_m(k) = \int \frac{d^3\epsilon}{(2\pi)^3}P_t(\vec{k}+\vec{\epsilon})W_s(\vec{\epsilon})
\label{eq:final}
\end{equation}

 Eq. \ref{eq:final} is expressed in terms of the theoretical spherically-averaged power spectrum (e.g., output by CAMB), and the spherically averaged window function. Thus we see that, because $P(k)$ is spherically symmetric, the expected value of the measured power only depends on the spherically averaged window function, which we estimate from the random catalog and denote $P_{win}$. Thus
\begin{equation}
P_m(k) = \int \int P_t(k+\epsilon) P_{win}(\epsilon)\epsilon^2  \delta(r_{\epsilon^{\prime}}-r_{\epsilon}){\rm d}{\epsilon}{\rm dcos}(\theta) ,
\label{eq:pmc}
\end{equation}
where  $r_{\epsilon^{\prime}} =  k+\epsilon$ and $r_{\epsilon} =  \sqrt{k^2+\epsilon^2-2k\epsilon {\rm cos}(\theta)}$. 

We measure the spherically averaged $P_m$ and $P_{win}$ in discrete $k$-bins. Further, determining \ref{eq:pmc} for every theoretical model tested would be computationally severe. To account for these facts, we fit a spline to $P_{win}$ and then determine the matrix $W[k_i][k_j]$ via
\begin{equation}
W[k_i][k_j] = \int \int P_{win}(\epsilon)\epsilon^2  \Theta(r_{\epsilon},k_j){\rm d}{\epsilon}{\rm dcos}(\theta)
\end{equation}
where now $r_{\epsilon} = \sqrt{k_i^2+\epsilon^2-2k_i\epsilon {\rm cos}(\theta)}$ and $\Theta(r_{\epsilon},k_j)$ is 1 if $r_{\epsilon}$ lies within the $k$-bin $k_j$ and 0 otherwise. Thus, we only need to calculate the input $P(k)$ at each bin $k_j$ and use Eq. \ref{eq:pkm} to determine the model for our measurement at $k_i$. Fig. \ref{fig:pkconv} and the surrounding text demonstrates that this methodology works extremely well on mock DR9 CMASS samples.

\section{Tests on Mocks}
\label{app:mocktest}

\begin{figure}
\includegraphics[width=84mm]{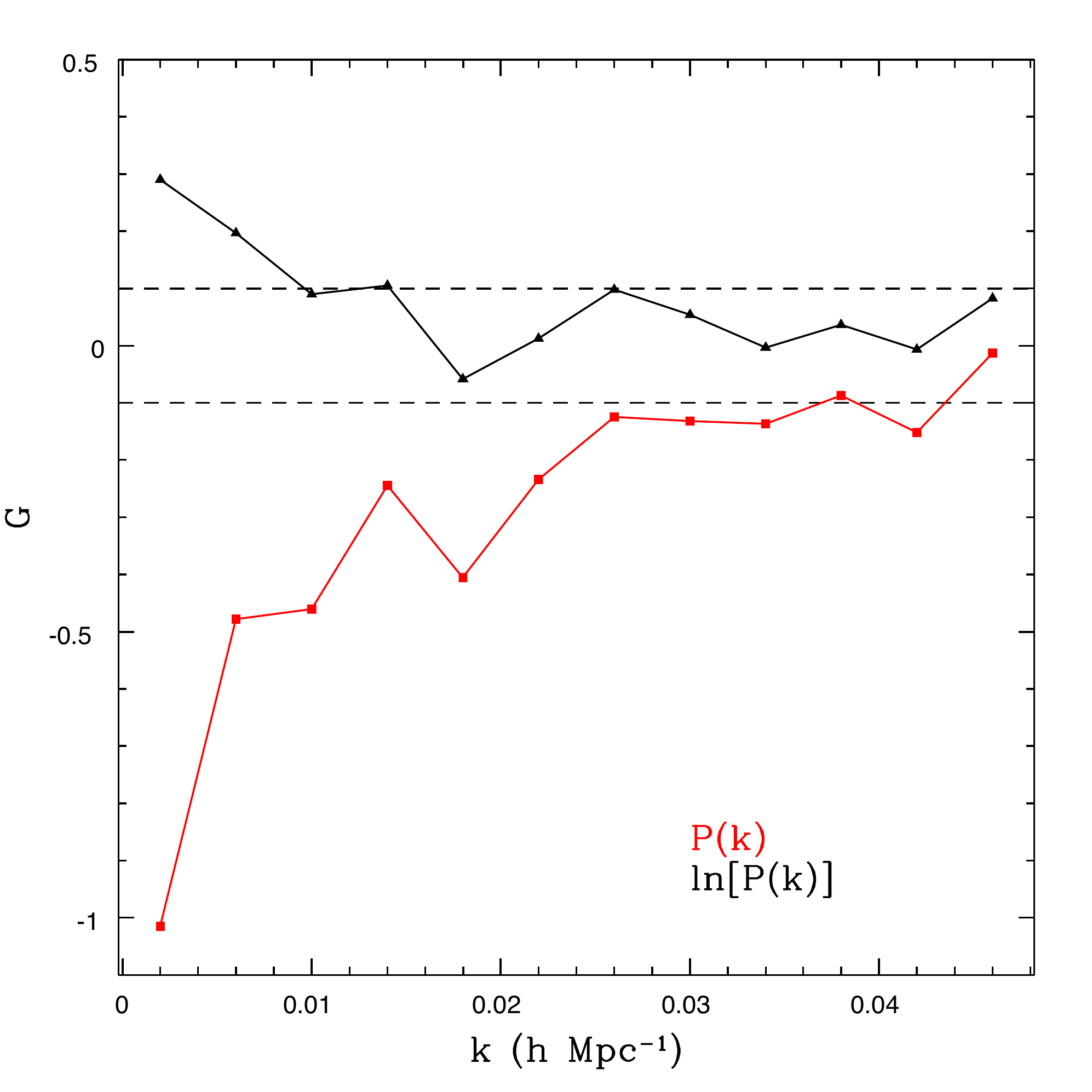}
  \caption{The skewness, $G$, of the distributions of the 600 mock power spectra (red) and of the logarithm of the 600 mock power spectra (black). The dashed lines display the expectation of the standard deviation of the skewness of 600 values drawn from a Gaussian distribution (9.975$\times 10^{-2}$).}
  \label{fig:skew}
\end{figure}

The 600 mock DR9 CMASS samples created by \cite{Manera12} allow us to test our methodology. We are using a covariance matrix constructed from the logarithms of the mock power spectra, $P(k)_{mock}$. This procedure is tested by evaluating the skewness, $G$, of the distributions of $P(k)_{mock}$ and ${\rm ln}[P(k)_{mock}]$ given by
\begin{equation}
G(k) = \frac{\sqrt{600\times599}}{600\times598}\sum^{i=600}_{i=1}\left(\frac{X(k)^i-\bar{X}(k)}{\sigma(k)}\right)^3,
\end{equation}
where $\sigma$ is the standard deviation, $X$ represents either $P(k)_{mock}$ or ${\rm ln}[P(k)_{mock}]$, and $\bar{X}$ represents the mean across all mocks. Fig. \ref{fig:skew} displays the skewness of the distributions of $P(k)_{mock}$ (red) and ${\rm ln}[P(k)_{mock}]$ (black) for the $k$-bins we consider in our analysis. The expectation of the standard deviation (9.975$\times 10^{-2}$) of the skewness of 600 values drawn from a Gaussian distribution is denoted by the black dashed line. For ${\rm ln}[P(k)_{mock}]$, three of the twelve skewness are greater than the expected standard deviation and only the skewness at the lowest $k$-bin is greater than two standard deviations from zero. For $P(k)_{mock}$, half of the skewness are greater than two standard deviations from zero. This suggests that testing ${\rm ln}[P(k)]$ against models using the covariance matrix constructed from ${\rm ln}[P(k)_{mock}]$ is indeed a better choice than $P(k)$ for the DR9 CMASS sample.

We test our treatment of systematics for cases (i), (iii), and (iv) on the power spectra determined from the mocks (the mocks have no systematic in their density field, so case (ii) is irrelevant). In general, we find that there is a degeneracy where in many case we find peaks in the $f_{\mathrm{NL}}^{\mathrm{local}}$ PDFs at both positive and negative values. This is due to the covariance: the best-fit value of $f_{\mathrm{NL}}^{\mathrm{local}}$ is most likely to occur when the model power spectrum has values in the two lowest-$k$ bins that are both either above or below the measured power spectrum. As summarized in Table \ref{tab:mocktest}, 9\% of the mocks in cases (i) and (iii) have PDFs with two peaks and 24\% of the PDFs for case (iv) have two peaks. We find that the mean $f_{\mathrm{NL}}^{\mathrm{local}}$ values (listed in Table \ref{tab:mocktest}) change by only {\bf0.6} when we test the different cases, suggesting that our treatment of systematics does not bias the recovered $f_{\mathrm{NL}}^{\mathrm{local}}$ value. 

While consistent for each case, the mean $f_{\mathrm{NL}}^{\mathrm{local}}$ values are $\sim$3$\sigma$ from the expected value of zero. This may be due to the fact that the distribution of ${\rm ln}[P(k)_{mock}]$ has a positive skewness (which implies median values smaller than the average) at the two lowest $k$-bins. This potential bias is much smaller than the uncertainty expected on any individual realization. Most importantly, the bias is consistent in each case and therefore should not alter any conclusions we reach on the effect of observational systematics on the measurement of $f_{\mathrm{NL}}^{\mathrm{local}}$.

\begin{table}
\caption{Statistics for tests on mock DR9 CMASS catalogs: frac. 68; 95 is the fraction of realizations that contain $f_{\mathrm{NL}}^{\mathrm{local}}=0$ within the area of the PDF containing the highest 68.2\%; 95\% of the likelihood, $\bar{f}_{\mathrm{NL}}^{\mathrm{local}}$ is the mean of the maximum likelihood value of $f_{\mathrm{NL}}^{\mathrm{local}}$, $\sigma$ is the standard deviation using $f_{\mathrm{NL}}^{\mathrm{local}} = 0$ as the mean, and frac. w. 2 peaks is the fraction of PDFs that have two peaks. }
\begin{tabular}{ccccc}
\hline
\hline
case  &  frac. 68; 95 &  $\bar{f}_{\mathrm{NL}}^{\mathrm{local}}$ & $\sigma$ & frac. w. 2 peaks  \\
\hline
i & 0.720$\pm$0.018; 0.960$\pm$0.008  & -7.0 & 57 & 0.090  \\
iii & 0.733$\pm$0.018; 0.957$\pm$0.008  & -7.0 & 57& 0.093 \\
iv & 0.688$\pm$0.018; 0.970$\pm$0.007  & -6.4 & 59& 0.240 \\
\hline
\label{tab:mocktest}
\end{tabular}
\end{table}

We further test our methods by determining the percentage of mocks realizations that find the correct answer ($f_{\mathrm{NL}}^{\mathrm{local}} = 0$) within 68.2\%/95\% (which will denote as the ``68/95 fraction'') of the area of the individual PDFs, i.e., we integrate the regions of the PDF with the greatest probability until we include $f_{\mathrm{NL}}^{\mathrm{local}} = 0$ and check if the integral is less than 0.682/0.95. We find that, given the preference for peaks at both positive and negative values, the best results are obtained when constructing the PDF of the absolute value of $f_{\mathrm{NL}}^{\mathrm{local}}$ when less than 90\% of the PDF is on one side of zero (i.e., $10\% < P(f_{\mathrm{NL}}^{\mathrm{local}}<0) < 90\%$). The results of this test are included in Table \ref{tab:mocktest}. We find that the 68 and 95 fractions are higher than we expect for all cases, but that the behavior of case (iv) differs from that of cases i and iii. For case (iv), the 68 fraction is within 1$\sigma$ of the expected value of 0.682, but the 95 fraction is 2.9$\sigma$ from the expected value of 0.95. In order to bring the 95 fraction determined for case (iv) within 1$\sigma$ of 0.95, we must reduce the integration threshold to greatest 92.3\% of the PDF. For both cases (i) and (iii), the 95 fractions are within 1.25$\sigma$ of the expected value, but the 68 fractions are greater than 2.1$\sigma$ from the expected value. In general, this suggests that the 95\% confidence intervals we determine are more robust than the 68.2\% confidence intervals, but that the width of the 95\% confidence interval determined for case (iv) may be over-estimated.

%


\section{Correlation Function}
\label{app:xi}
Our study focuses on the results that can be obtained from measurements of the power-spectrum of DR9 CMASS galaxies. Since the correlation function and power spectrum contain the same information, in principle, the same results should be achievable from measurements of the correlation function as our power spectrum results. Here we present the analysis we performed on the correlation function.

The spherically-averaged redshift space galaxy correlation function, $\xi_o(s)$, is the Fourier transform of $P^o_g(k)$. Inspection of Eqs. \ref{eq:Pexp} and \ref{eq:Pgo} reveals, that, for a given background cosmology and redshift, four integrals over $k$ are required to obtain $\xi_o(s,b,f_{\mathrm{NL}}^{\mathrm{local}})$. However, for the model $\xi_o(s)$, we must also account for the non-linear smoothing of the baryon acoustic oscillation (BAO) feature, as the damping of high $k$ BAO oscillations damps the (one) $\xi(s)$ BAO peak. We therefore use a damped matter power spectrum, $P_{Md}(k)$, obtained using a damping scale $s_d$ (see, e.g., \citealt{Seo05,CS06}) and:
\begin{equation}
P_{Md}(k) = P_M(k)e^{-\left[ks_d\right]^2}
\end{equation}

 Given $P_{Md}(k)$ and $P_M(k)$, we Fourier transform to determine the damped and undamped isotropic 3-dimensional real-space correlation functions $\xi_{lin,d}(r)$ and $\xi_{lin}$. We also include the effects of coupling between low and high $k$ modes (see, e.g., \citealt{crocce08,Sanchez08}) via
\begin{equation}
\xi(r) = \xi_{lin,d}(r)+A_{mc}\xi^{(1)}_{lin}(r)\xi^{\prime}_{lin}(r),
\end{equation}
where $\xi^{\prime}_{lin}$ is the derivative of $\xi_{lin}$ and
\begin{equation}
\xi_{lin}^{(1)} \equiv \frac{1}{2\pi^2} \int P_{M}(k)j_1(kr)kdk.
\end{equation}
Given $\xi(r)$, one can calculate the correlation function terms analogous to the power spectrum terms in Eq. \ref{eq:Pgo}. 

\begin{figure}
\includegraphics[width=84mm]{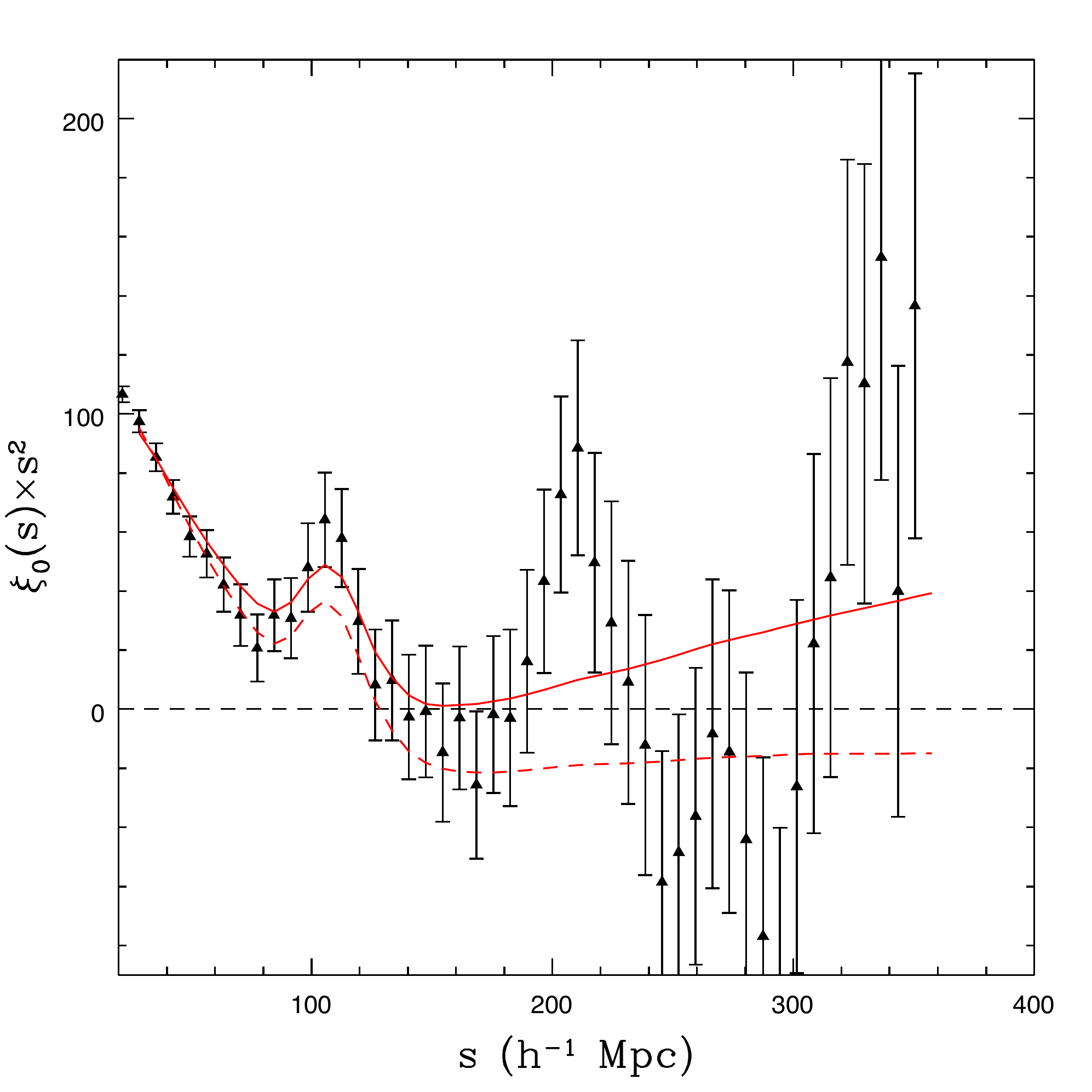}
  \caption{The measured correlation function of the DR9 CMASS sample, multiplied by $s^2$ (black) and the best-fit model, with $f_{\mathrm{NL}}^{\mathrm{local}} =$ {\bf107}, shown by the solid red line. The black dashed line denotes $\xi(s)=0$ and the dashed red line displays a model with $f_{\mathrm{NL}}^{\mathrm{local}} = 0$. The baryon acoustic oscillation feature is seen at $\sim100h^{-1}$Mpc. The significance of the apparent peak at $\sim200h^{-1}$Mpc is studied in Ross et al (2012). }
  \label{fig:xibf}
\end{figure}

\begin{figure}
\includegraphics[width=84mm]{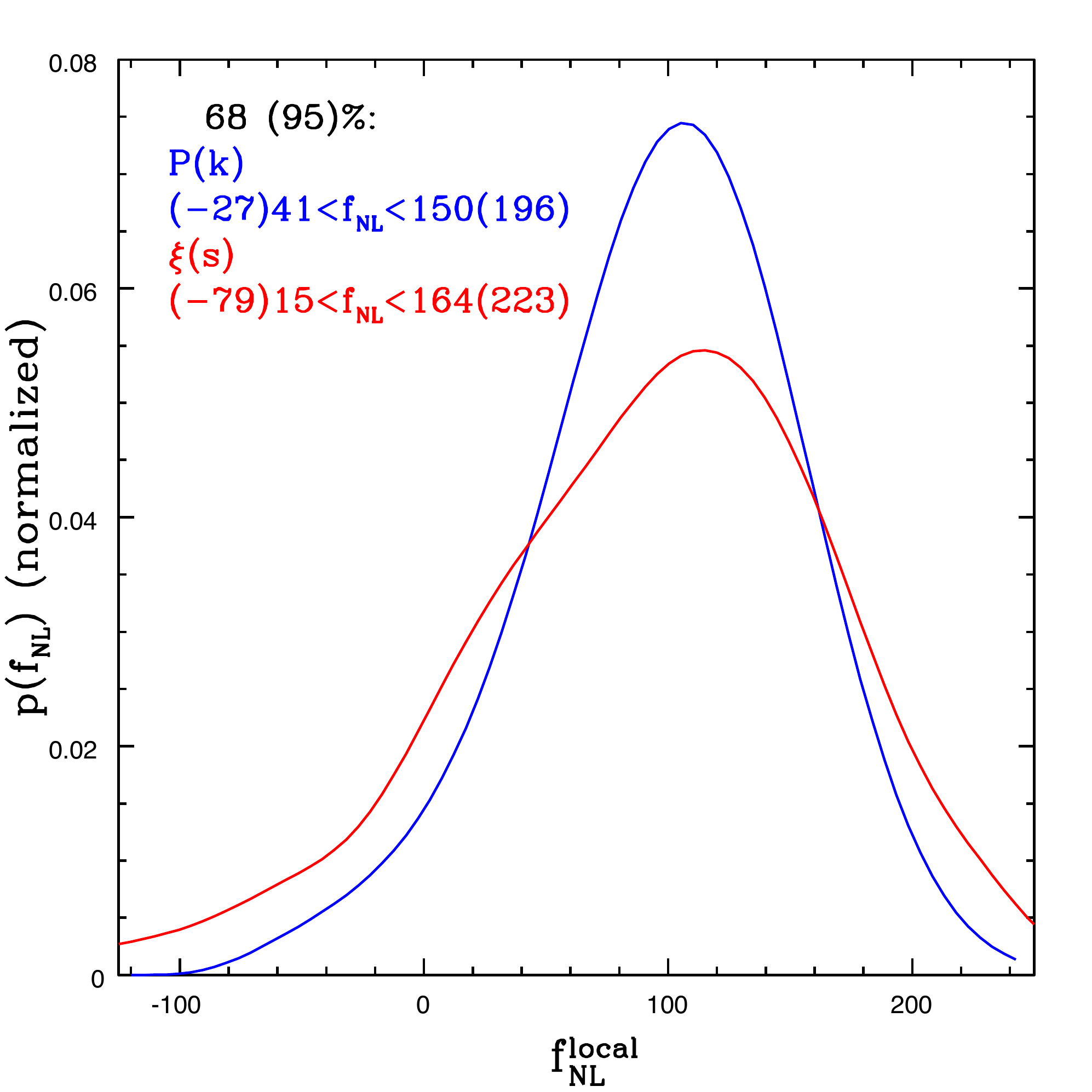}
  \caption{The normalized (so that they integrate to 1) probability distributions for the local non-Gaussianity parameter $f_{\mathrm{NL}}^{\mathrm{local}}$, obtained from the power-spectrum ($P(k)$; blue) and the correlation function ($\xi(s)$; red), both using the fiducial systematic correction applied to the DR9 CMASS sample. The distributions have similar maximum likelihood values for $f_{\mathrm{NL}}^{\mathrm{local}}$, but the $\xi(s)$ measurements allow a wider range of $f_{\mathrm{NL}}^{\mathrm{local}}$ values. }
  \label{fig:pfnlxip}
\end{figure}

We measure the correlation function, $\xi_{meas}(s)$ for the observed DR9 CMASS sample using the standard \cite{LZ} estimator given by
\begin{equation}
\xi_{meas}(s) = \frac{DD(s)-2DR(s)}{RR(s)}+1,
\end{equation}
\noindent where $D$ represents the data sample (i.e., BOSS galaxies) and $R$ represents the random sample (occupying the angular footprint and with the same redshift distribution as the data sample) and the paircounts are normalized to the total number. We use a linear spaced binning in $s$ of 7 $h^{-1}$Mpc, and the measurements are the same as those presented in \cite{Ross12}.

By definition, the correlation function integrates to zero:
\begin{equation}
\int_0^{\infty} \xi(s)ds = 0
\end{equation}
When estimating the correlation function, this condition holds, but we do not have an infinite volume. Instead the condition becomes a sum over the pairs from which $\xi_{meas}$ is determined
\begin{equation}
\frac{\sum_{i} RR(s_i)\xi(s_i)_{meas}}{\sum RR(s_i)} = 0.
\end{equation}
Thus, for any particular model $\xi(s)_{mod}$, we can compute the deviation from this condition as 
\begin{equation}
I_{mod} = -\frac{\sum RR(s)\xi(s)_{mod}}{\sum RR(s)},
\end{equation}
and thus the total model to compare to the measurement is $\xi(s)_{mod} + I_{mod}$. For our fiducial model, and a bias of 1.9, it is $I_{mod} = 5.6\times10^{-5}$, but for $f_{\mathrm{NL}}^{\mathrm{local}} =$ 50, it is $I_{mod} = 1.3\times10^{-4}$ (for reference, the standard deviation of the $\xi(s)$ computed from the mock catalogs is an order of magnitude larger, e.g., it is 0.002 at $s=200h^{-1}$Mpc).

Using the correlation function (as opposed to the power power spectrum) measurements presents additional challenges for the covariance matrix that is applied. The amplitude of the measurements are generally expected to cross zero just beyond the BAO scale, and where they are close to zero we should not expect the fractional error to be conserved for alternative models. We thus use $\xi(s)$ and its covariance, as determined from the 600 mocks, accepting that this may be inaccurate for large $f_{\mathrm{NL}}^{\mathrm{local}}$, but that the results will still be useful for comparison. Further, the model $\xi(s)$ has two nuisance parameters, $s_{d}$ and $A_{mc}$, that the power spectrum does not have. For the results we present, we marginalize over these parameters, as in \cite{SanchezCos12}, but we find less than a 10\% variation in the width and maximum of likelihood distribution when we fix the damping scale to the value used in \cite{alph}.  

Fig. \ref{fig:xibf} displays the measured $\xi_0$ for the DR9 CMASS sample and the best-fit model $\xi_0$, applying case (i), for which we find $f_{\mathrm{NL}}^{\mathrm{local}}=$ 107, consistent with the value of $f_{\mathrm{NL}}^{\mathrm{local}}=$ 105 we found using the power spectrum. Fig. \ref{fig:pfnlxip} displays the probability distribution we find for $f_{\mathrm{NL}}^{\mathrm{local}}$ using the $\xi_o(s)$ measurement with $30 < s < 400 h^{-1}$Mpc (red) and the $P_o(k)$ measurements at $k < 0.05h$Mpc$^{-1}$ (both for the measurements using the fiducial case i). Each prefer positive $f_{\mathrm{NL}}^{\mathrm{local}}$ at greater than 68\% confidence, but allow $f_{\mathrm{NL}}^{\mathrm{local}} < 0$ at greater than 5\%. As expected, the two estimators recover consistent results. However, the probability distribution is considerably wider for the correlation function measurements. This is likely due to the fact that the maximum effective scale probed is much larger for the power-spectrum measurements, for which the data compression at large scales is more effective. This result justifies our choice to focus on the power-spectrum results.

\end{appendix}
\label{lastpage}

\end{document}